\documentclass[aps,prl,twocolumn,superscriptaddress,nobibnotes]{revtex4-1}

\setcitestyle{super} %% set superscript for reference numbers
\usepackage{amsmath,amssymb}
\usepackage{graphicx}
\usepackage{color} 
\usepackage[T1]{fontenc}

\usepackage{xspace}  
\usepackage{multirow}
\usepackage{dcolumn} %centers to the decimal point in tables using ``d''
\usepackage{booktabs}
\usepackage{ulem}

\renewcommand{\figurename}{\textbf{Fig.}}
\renewcommand{\tablename}{\textbf{Table}}

\usepackage{caption}
\DeclareCaptionLabelFormat{nature}{\textbf{#1\ #2}\ $\vert$\ }
\usepackage{xspace}  
\newcommand{\ts}{\textsuperscript} 
\newcommand{\etal}{\textit{et al.}\xspace}
\newcommand{\gammaray}{\ensuremath{\gamma}-ray\xspace}
\newcommand{\gammarays}{\ensuremath{\gamma} rays\xspace} %% shold be noun
\newcommand{\mevu}{MeV/$u$\xspace}

\newcommand{\etwop}{\ensuremath{E(2^+_1)}\xspace}
\newcommand{\efourp}{\ensuremath{E(4^+_1)}\xspace}
\newcommand{\stwop}{\ensuremath{2^+_1}\xspace}

\newcommand{\eratior}{\ensuremath{R_{4/2}}\xspace}
\newcommand{\etwopt}{\ensuremath{2^+_1 \rightarrow 0^{+}_{\mathrm{gs}}}\xspace}
\newcommand{\efourpt}{\ensuremath{4^+_1 \rightarrow 2^+_1}\xspace}
\begin{document}

%\preprint{}

%\title{First spectroscopy of \ts{78}Ni:\\ At the frontier of a new paradigm for nuclear structure}
%\title{\ts{78}Ni: a doubly magic nucleus on the shore of a new island of inversion}
%\title{Doubly Magic \ts{78}Ni: A Stronghold against Nuclear Deformation}
\title{\ts{78}Ni revealed as a doubly magic stronghold against nuclear deformation} % no punctuation

\newcommand{\acea}{        \affiliation{IRFU, CEA, Universit\'e Paris-Saclay, F-91191 Gif-sur-Yvette, France}}
\newcommand{\ariken}{      \affiliation{RIKEN Nishina Center, 2-1 Hirosawa, Wako, Saitama 351-0198, Japan}}
\newcommand{\atudarmstadt}{\affiliation{Institut f\"ur Kernphysik, Technische Universit\"at Darmstadt, 64289 Darmstadt, Germany}}
\newcommand{\aiphc}{       \affiliation{IPHC, CNRS/IN2P3, Universit\'e de Strasbourg, F-67037 Strasbourg, France}}
\newcommand{\acns}{        \affiliation{Center for Nuclear Study, The University of Tokyo, RIKEN campus, Wako, Saitama 351-0198, Japan}}
\newcommand{\aut}{         \affiliation{Department of Physics, The University of Tokyo, 7-3-1 Hongo, Bunkyo, Tokyo 113-0033, Japan}}
\newcommand{\arikkyo}{     \affiliation{Department of Physics, Rikkyo University, 3-34-1 Nishi-Ikebukuro, Toshima, Tokyo 172-8501, Japan}}
\newcommand{\abeijing}{    \affiliation{State Key Laboratory of Nuclear Physics and Technology, Peking University, Beijing 100871, P.R. China}}
\newcommand{\abrighton}{   \affiliation{School of Computing, Engineering and Mathematics, University of Brighton, Brighton BN2 4GJ, United Kingdom}}
\newcommand{\ainst}{       \affiliation{Institute for Nuclear Science \& Technology, VINATOM, P.O. Box 5T-160, Nghia Do, Hanoi, Vietnam}}
\newcommand{\aatomki}{     \affiliation{MTA Atomki, P.O. Box 51, Debrecen H-4001, Hungary}}
\newcommand{\aipno}{       \affiliation{Institut de Physique Nucl\'eaire, CNRS/IN2P3, Universit\'e Paris-Saclay, F-91406 Orsay Cedex, France}}
\newcommand{\aoslo}{       \affiliation{Department of Physics, University of Oslo, N-0316 Oslo, Norway}}
\newcommand{\ahku}{        \affiliation{Department of Physics, The University of Hong Kong, Pokfulam, Hong Kong}}
\newcommand{\asurrey}{     \affiliation{Department of Physics, University of Surrey, Guildford GU2 7XH, United Kingdom}}
\newcommand{\atohoku}{     \affiliation{Department of Physics, Tohoku University, Sendai 980-8578, Japan}}
\newcommand{\aarpajon}{    \affiliation{CEA, DAM, DIF, 91297 Arpajon, France}}
\newcommand{\arcnp}{       \affiliation{Research Center for Nuclear Physics, Osaka University, Ibaraki 567-0047, Japan}}
\newcommand{\aocu}{	   \affiliation{Department of Physics, Osaka City University, Osaka 558-8585, Japan}}
\newcommand{\atriumf}{     \affiliation{TRIUMF 4004 Wesbrook Mall, Vancouver, British Columbia, Canada V6T 2A3}}
\newcommand{\auam}{        \affiliation{Departamento de F\'isica Teorica and IFT-UAM/CSIC, Universidad Aut\'onoma de Madrid, 28049 Madrid, Spain}}
\newcommand{\aemmi}{       \affiliation{ExtreMe Matter Institute EMMI, GSI Helmholtzzentrum f\"ur Schwerionenforschung GmbH, 64291 Darmstadt, Germany}}
\newcommand{\aheidelberg}{ \affiliation{Max-Planck-Institut f\"ur Kernphysik, Saupfercheckweg 1, 69117 Heidelberg, Germany}}
\newcommand{\amainz}{      \affiliation{Institut f\"ur Kernphysik and PRISMA Cluster of Excellence, Johannes Gutenberg-Universit\"at Mainz, 55128 Mainz, Germany}}
\newcommand{\areed}{       \affiliation{Physics Department, Reed College, 3203 SE Woodstock Blvd., Portland, OR 97202-8199, USA}}
\newcommand{\acsnsm}{       \affiliation{CSNSM, CNRS/IN2P3, Universit\'e  Paris-Saclay, F-91405 Orsay Campus, France}}

\newcommand{\aemp}{\email{pieter@ribf.riken.jp}}  % author email
\newcommand{\aema}{\email{alexandre.obertelli@cea.fr}}  % author email
\newcommand{\aemt}{\email{taniuchi@ribf.riken.jp}}  % author email
 
\author{R.~Taniuchi}     \aut \ariken %\aemt
\author{C.~Santamaria}   \acea \ariken  
\author{P.~Doornenbal}   \aemp \ariken 
\author{A.~Obertelli}    \acea \ariken  \atudarmstadt
\author{K.~Yoneda}       \ariken
%%Core team members
\author{G.~Authelet}     \acea
\author{H.~Baba}         \ariken
\author{D.~Calvet}       \acea
\author{F.~Ch\^ateau}    \acea
\author{A.~Corsi}        \acea
\author{A.~Delbart}      \acea
\author{J.-M.~Gheller}   \acea
\author{A.~Gillibert}    \acea
\author{J.D. Holt}       \atriumf 
\author{T.~Isobe}        \ariken
\author{V.~Lapoux}       \acea
\author{M.~Matsushita}   \acns        
\author{J.~Men\'endez}   \acns
\author{S.~Momiyama}     \aut \ariken 
\author{T.~Motobayashi}  \ariken      
\author{M.~Niikura}      \aut
\author{F.~Nowacki}      \aiphc
\author{K.~Ogata}        \arcnp \aocu
\author{H.~Otsu}         \ariken
\author{T.~Otsuka}       \acns \aut \ariken
\author{C.~P\'eron}      \acea
\author{S.~P\'eru}       \aarpajon
\author{A.~Peyaud}       \acea
\author{E.C.~Pollacco}   \acea
\author{A. Poves}        \auam
\author{J.-Y.~Rouss\'e}  \acea
\author{H.~Sakurai}      \aut \ariken 
\author{A.~Schwenk}      \atudarmstadt \aemmi \aheidelberg
\author{Y.~Shiga}        \ariken \arikkyo
\author{J.~Simonis}      \amainz \atudarmstadt \aemmi
\author{S.R.~Stroberg}   \atriumf \areed
\author{S.~Takeuchi}     \ariken       
\author{Y.~Tsunoda }     \acns  
\author{T.~Uesaka}       \ariken      
\author{H.~Wang}         \ariken      
%Non core team members
\author{F.~Browne}       \abrighton   
\author{L.X.~Chung}      \ainst
\author{Zs.~Dombradi}    \aatomki
\author{S.~Franchoo}     \aipno
\author{F.~Giacoppo}     \aoslo
\author{A.~Gottardo}     \aipno
\author{K.~Hady\'nska-Kl\c{e}k}\aoslo
\author{Z.~Korkulu}      \aatomki
\author{S.~Koyama}       \aut \ariken 
\author{Y.~Kubota}       \ariken \acns
\author{J.~Lee}          \ahku
\author{M.~Lettmann}     \atudarmstadt
\author{C.~Louchart}     \atudarmstadt
\author{R.~Lozeva}       \aiphc \acsnsm
\author{K.~Matsui}       \aut \ariken 
\author{T.~Miyazaki}     \aut \ariken 
\author{S.~Nishimura}    \ariken      
\author{L.~Olivier}      \aipno
\author{S.~Ota}          \acns
\author{Z.~Patel}        \asurrey
\author{E.~\c{S}ahin}        \aoslo
\author{C.~Shand}        \asurrey
\author{P.-A.~S\"oderstr\"om}\ariken
\author{I.~Stefan}       \aipno
\author{D.~Steppenbeck}  \acns
\author{T.~Sumikama}     \atohoku
\author{D.~Suzuki}       \aipno
\author{Zs.~Vajta}       \aatomki
\author{V.~Werner}       \atudarmstadt
\author{J.~Wu}           \ariken \abeijing  
\author{Z.Y.~Xu}           \ahku 

%\date{\today}
\date{February 10, 2019}
\begin{abstract}
\bf{
%wordcount 200 -> Upto 150 words for article
  Nuclear magic numbers, which emerge from the strong nuclear force based on quantum chromodynamics, 
  correspond to fully occupied energy shells of protons, or neutrons inside atomic nuclei. 
  Doubly magic nuclei, with magic numbers for both protons and neutrons, are spherical and extremely rare across the nuclear landscape. 
  While the sequence of magic numbers is well established for  
  stable nuclei, evidence reveals modifications for nuclei with a large proton-to-neutron asymmetry.
  Here, we provide the first spectroscopic study of the doubly magic nucleus \ts{78}Ni, fourteen neutrons beyond the last stable
  nickel isotope.  
  We {provide direct evidence for} its doubly magic nature, which is also predicted by \textit{ab initio} 
  calculations based on chiral effective field theory interactions and the quasi-particle random-phase approximation.
  However, our results also provide the first indication of the
  breakdown of the neutron magic number 50 and proton magic number 28 beyond this stronghold, caused by a competing deformed 
  structure. State-of-the-art phenomenological shell-model calculations reproduce this shape coexistence, predicting further a rapid
  transition from spherical to deformed ground states with \ts{78}Ni as turning point.
} 
\end{abstract}

\maketitle
%%%%%%%%%%%%%%%
% Fig1: E2 systematics.
\begin{figure*}
  \centering
  \includegraphics[width=18.3cm, keepaspectratio]{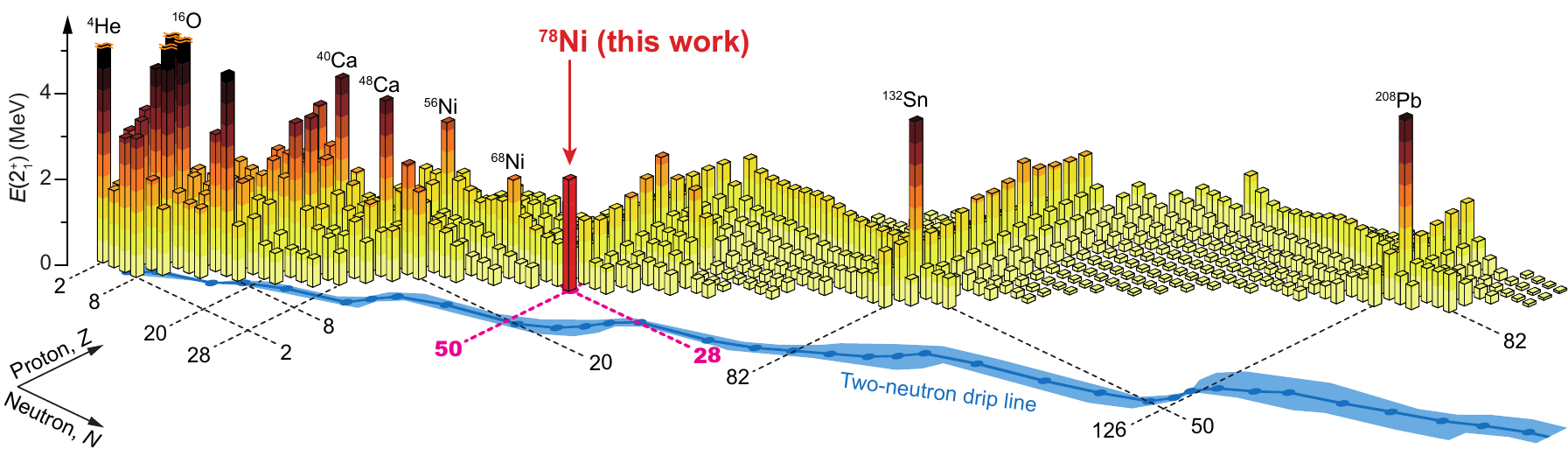}%[width=18.3cm]
  \caption{{\bf Experimental \mbox{\boldmath{\etwop}} systematics of the even-even nuclear landscape}. Shown are known \etwop of even-even isotopes~\cite{ensdf}
    and the value for \ts{78}Ni obtained in the present study. 
    Traditional magic numbers are indicated by dashed lines and doubly magic nuclei are labelled.
    Also \ts{68}Ni, for which the number of neutrons, $N=40$, matches the harmonic oscillator shell closure, is marked.  
    The predicted two-neutron drip line and its uncertainties~\cite{erler:2012:NATURE} are shown in blue.
  } 
  \label{fig:e2systematics}  
\end{figure*}  

% Fig2: Layout of Facility, MINOS, and PID spectra.
\begin{figure}
  \centering
  \includegraphics[scale=1.0, keepaspectratio]{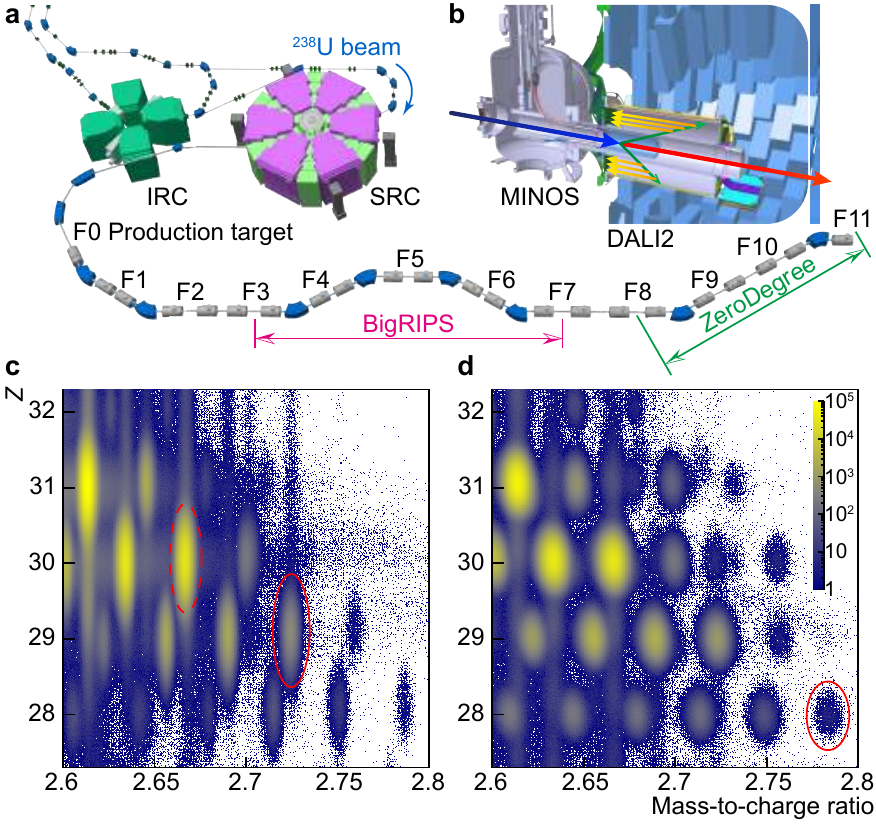}%[width=8.9cm]
  \caption{{\bf Layout of the experimental equipment and particle identification plots of isotopes}.
    \textbf{a}, Schematic view of the final two cyclotron stages, IRC and SRC, of the RIBF facility
    along with the BigRIPS and ZeroDegree fragment separators. Reaction residues cover a flight length of 118~m between creation at the F0 production target 
    and the final focal point, F11.
    \textbf{b}, A layout of MINOS and the surrounding DALI2 \gammaray spectrometer located at the F8 focal plane. % 'located ~' added    is presented. 
    \textbf{c}, Components of the radioactive beam accepted by the BigRIPS fragment separator. Events corresponding to \ts{79}Cu (red ellipse) and \ts{80}Zn 
    (dashed red ellipse) are indicated.
    \textbf{d}, Reaction products accepted by ZeroDegree. \ts{78}Ni is enclosed by the red ellipse.
    Both plots in \textbf{c}, \textbf{d} share the colour scale that indicates the number of events for the different isotopes transmitted through 
    BigRIPS and ZeroDegree.
  }   
  \label{fig:pid}  
\end{figure}  

% Fig3: Experimental spectroscopy results from p2p and p3p.
%%%%%%%%%%%%%%%%%%%%%%%%%%%%%%%  
\begin{figure}
  \centering 
  \includegraphics[width=8.9cm, keepaspectratio]{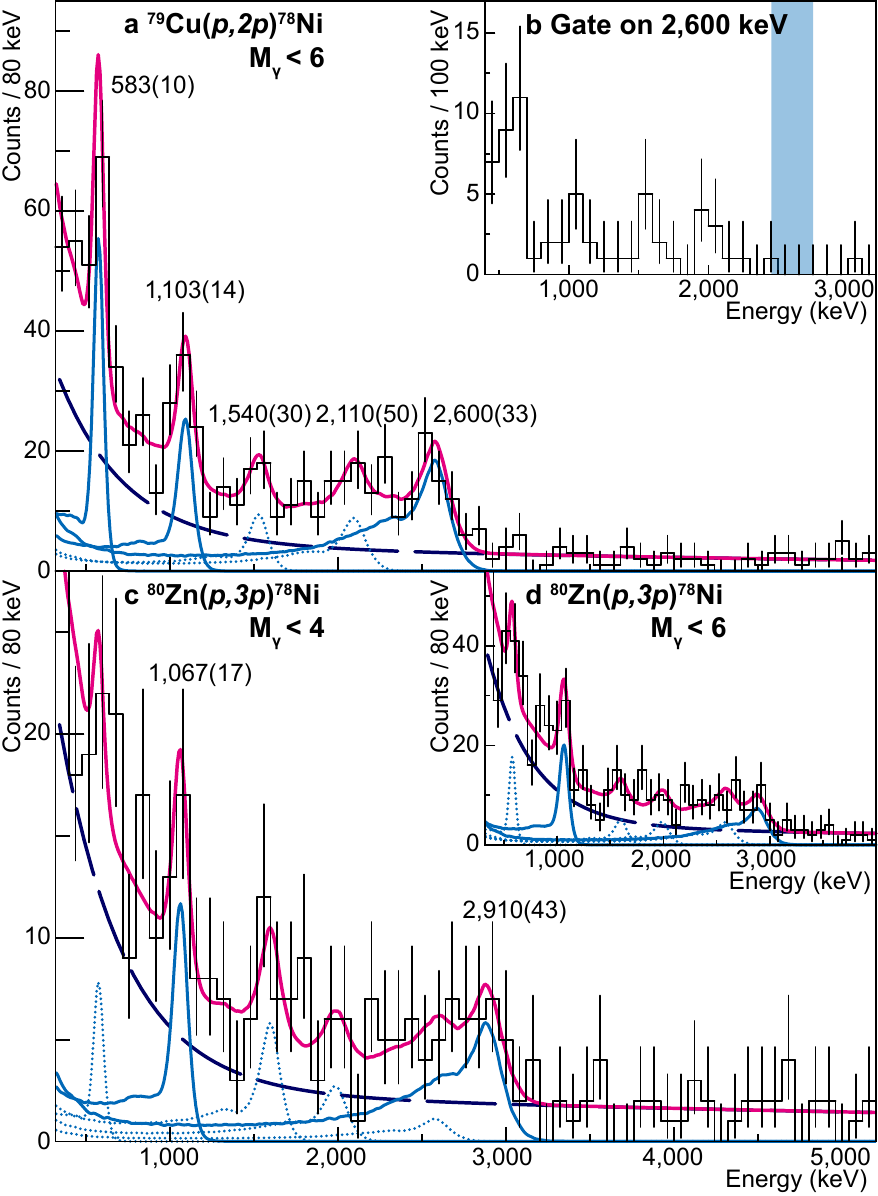}
  \caption{{\bf Doppler-corrected \gammaray energy spectra}.
  \textbf{a}, De-excitation \gammarays measured in coincidence with \ts{78}Ni following $(p,2p)$ reactions. 
  \textbf{b}, Prompt coincidences with the 2,600-keV transition following $(p,2p)$ reactions. 
    It reveals coincidences between the 2,600-keV and other low-lying transitions. 
    The coincidence range is illustrated by the hatched area. 
%% p3p
  \textbf{c}, De-excitation \gammarays measured in coincidence with \ts{78}Ni following $(p,3p)$ reactions.
    To reduce events with multiple \gammaray hits from Compton scattering, the \gammaray detection multiplicity, $M_{\gamma}$, was restricted to values below 
    four. As a consequence, the visibility of the 2,910-keV transition is enhanced.
  \textbf{d}, Same as \textbf{c}, but for the $M_{\gamma}<6$ condition. 
  Observed transitions are indicated by their energies in \textbf{a} and \textbf{c}.
  Simulated response curves of the \gammaray detector for the individual transitions are illustrated by the light-blue solid ($\textrm{S.L.}\geq3$) and 
  dotted ($\textrm{S.L.}<3$) lines, while the fitted double-exponential background is shown by the dark-blue dashed line. Background and individual transitions 
  are summed for the magenta solid lines. Vertical error bars correspond to 1~s.d. errors. 
  } 
\label{fig:gammaspectrum}
\end{figure}  
%%%%%%%%%%%%%%%%% 

%%%%%%%%%%%%%%%%%  
\begin{figure*}  
  \centering
  \includegraphics[width=17.3cm]{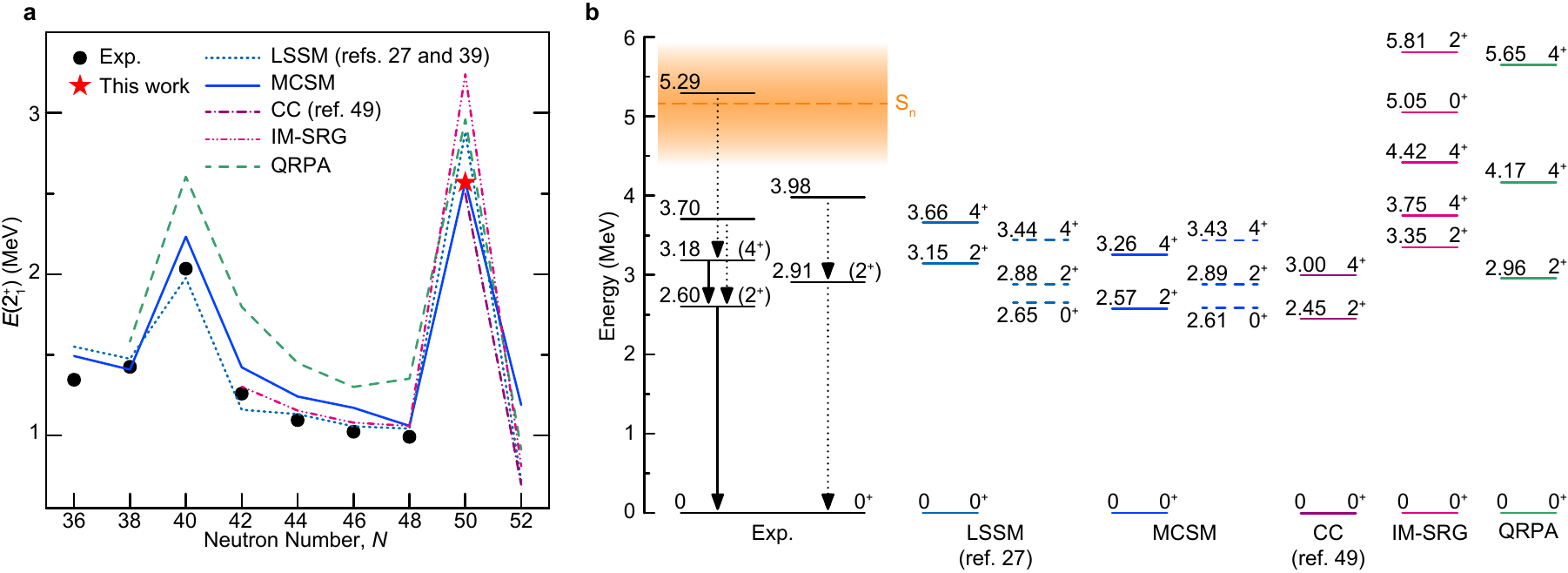}
  \caption{{\bf Comparison of theoretical predictions with experimental data}.
    \textbf{a}, Experimental \etwop for even-even nickel isotopes are compared to calculations using phenomenological shell-model 
    interactions, LSSM~\cite{nowacki:2016:PRL,lenzi:2010:PRC} (dotted line) and MCSM~\cite{tsunoda:2014:PRC} (solid line), 
    the beyond mean field approach, QRPA (dashed line), and the {\it ab initio} approach, 
    IM-SRG (dash-dot-dotted line) and CC~\cite{hagen:2016:PRL} (dash-dotted line), as a function of their neutron number, $N$. 
    The present result for \ts{78}Ni is indicated by the red star.
    \textbf{b}, Deduced experimental level scheme compared to detailed theoretical calculations for \ts{78}Ni.
    Transitions with S.L. $\ge 5$ are represented by solid arrows. Dotted arrows correspond to values $\mathrm{S.L.}<5$.
    The 1.54-MeV transition is not placed, {while the 2.11-MeV transition has a $\mathrm{S.L.}<3$.} 
    The evaluated neutron separation energy, $S_{n}$, and its errors~\cite{wang:2017:CPC} are indicated 
    by the orange dashed line and area, respectively. Predicted deformed states are indicated by dashed lines. 
    For convenience, theoretical predictions show only the two lowest $0^+,\ 2^+$, and $4^+$ spin-parity values.
}%    
\label{fig:levelscheme}% 
\end{figure*}  
%%%%%%%%%%%%%%%%%

Characterisation of the few doubly magic nuclei, known and predicted, provides a benchmark for our knowledge of the fundamental forces that drive
the evolution of shell closures with proton-to-neutron asymmetry~\cite{wienholtz:2013:NATURE,steppenbeck:2013:NATURE}. With reliable and
globally applicable interactions, accurate predictions of the location of the two-neutron drip line and the nuclear landscape can be made~\cite{erler:2012:NATURE}.
These, in turn, are critical for nucleosynthesis models, which rely on nuclear structure inputs.

An initial characterisation of magicity is often provided by the first $J^{\pi}= 2^{+}$ excitation energy, \etwop, as illustrated in
Fig.~\ref{fig:e2systematics} for the Segr\`e chart, a two-dimensional grid in which nuclei are arranged by their 
proton ($Z$) and neutron ($N$) numbers. Magic nucleon numbers (2, 8, 20, 28, 50, 82, 126), which were first correctly reproduced theoretically
for stable isotopes by introducing a strong spin-orbit interaction~\cite{jensen:1949:PR,mayer:1949:PR}, stand out, 
as excitation from the ground state requires promoting nucleons across major nuclear shells, and, therefore, more energy due to the large energy 
gaps involved. {Further experimental observables, such as charge radii or reduced $\gamma$-ray transition probabilities, are
indispensable for a comprehensive characterisation  of a nucleus. Their acquisition is, however, experimentally challenging.}

With the extension of studies to unstable, radioactive isotopes (RIs) with a large neutron excess -- also termed `exotic' nuclei --,
magic numbers emerged as a local feature. In lieu, nuclear shell structure changes, sometimes drastically, with 
the number of protons and neutrons, revealing interesting properties of the underlying nuclear forces. For instance, it was recognised
that several traditional neutron magic numbers disappear far from stability, such as 
$N=8, 20, 28$~\cite{navin:2000:PRL,thibault:1975:PRC,guillemaud-mueller:1984:NPA,bastin:2007:PRL}, while new ones have been claimed at 
$N=16$~\cite{tshoo:2012:PRL} and $N=32, 34$~\cite{huck:1985:PRC,wienholtz:2013:NATURE,steppenbeck:2013:NATURE}.

Shifts of these magic numbers challenge nuclear theory, and certain cases can be explained by 
drifts of the single-particle orbitals (SPOs) with varying nucleon number~\cite{talmi:1960:PRL}.
The central potential of the nucleon-nucleon ($NN$) effective interaction and the tensor force contribute strongly to this 
evolution~\cite{otsuka:2005:PRL,otsuka:2010:PRLa}. Also three-nucleon (3$N$) forces, which originate from the composite nature of nucleons, 
have a significant impact~\cite{otsuka:2010:PRL,hammer:rmp:2013}. So far, a coherent picture of the nuclear shell structure and
its evolution towards the most neutron-rich nuclei remains elusive.

The isotope \ts{78}Ni (28 protons and 50 neutrons) provides a unique case included in all motivations for planned and operating next-generation 
RI beam in-flight facilities, such as the RIBF in Japan, FRIB in the USA, and FAIR in Germany. 
Predictions regarding the neutron drip-line location~\cite{erler:2012:NATURE} of even-even nuclei, for which the two-neutron separation energy becomes
negative (also shown in Fig.~\ref{fig:e2systematics}), reveal that, prior to the measurement reported here, \ts{78}Ni was the only neutron-rich doubly 
magic nucleus lacking spectroscopic information on excited states that can be reached with current and next-generation facilities. 

%%%
After the first production of \ts{78}Ni~\cite{engelmann:1995:ZPA}, enormous efforts have been put into investigating its structure.
Previous measurements indirectly inferred persistent $N=50$~\cite{walle:2007:PRL,hakala:2008:PRL,mazzocchi:2000:PLB,hosmer:2005:PRL,xu:2014:PRL} 
and $Z=28$~\cite{olivier:2017:PRL,sahin:2017:PRL,welker:2017:PRL} shell closures at \ts{78}Ni. {This notion has been reinforced 
theoretically by \textit{ab initio} predictions~\cite{hagen:2016:PRL}.}  
Conversely, studies of \ts{66}Cr ($Z=24$) and \ts{70,72}Fe ($Z=26$) revealed constantly low excitation energies that question the $N=50$ shell closure 
for proton numbers below $Z=28$\cite{santamaria:2015:PRL, nowacki:2016:PRL}. Likewise, several studies supported a reduction of the $Z=28$ proton shell 
gap towards and beyond $N=50$\cite{franchoo:1998:PRL, flanagan:2009:PRL, sieja:2010:PRC, gottardo:2016:PRL, yang:2016:PRL}.
{A vanishing of the proton and neutron shell closures would be accompanied by an onset of deformation, implying dramatic 
consequences: Shape coexistence and gain in nuclear binding energy. The former signifies occurrence of several quantum states of different shapes
lying close and low in energy, the latter slants the two-neutron drip line and accordingly the limits of nuclear existence towards heavier isotopes.}
Hitherto, no ultimate conclusion on the magic character of \ts{78}Ni existed.
Here, we provide first direct evidence from in-beam $\gamma$-ray spectroscopy in prompt coincidence with proton removal reactions of fast RI beams. 

\section{P\lowercase{roduction of} RI \lowercase{beams}} % 35/40 letters

The experiment was carried out at %a world-class heavy-ion accelerator complex, the Radioactive Isotope Beam Factory (RIBF), Japan
the RIBF, which combines three injectors with four coupled cyclotrons.
Neutron-rich RI beams were produced by induced relativistic in-flight fission of a 345~\mevu \ts{238}U primary beam on a 3-mm-thick beryllium production target,
located at the F0 focus of the BigRIPS fragment separator~\cite{kubo:2012:PTEP} shown in Fig.~\ref{fig:pid}a. 
\ts{79}Cu and \ts{80}Zn particles, produced at rates of 5 and 290 particles per second, respectively, %via relativistic in-flight fission, 
were identified on an event-by-event basis from focal plane F3 to F7, before being guided to the MINOS reaction target system~\cite{obertelli:2014:EPJA} 
(see Fig.~\ref{fig:pid}b) located at F8 with a remaining energy of approximately 250~\mevu (61\% of the speed of light).

\section{\lowercase{\gammaray detection after secondary reaction}} % 40/40 letters

MINOS was composed of a 102(1)-mm-thick liquid hydrogen target and a time projection chamber to reconstruct the 
reaction vertices. This allowed to overcome inaccuracies in the Doppler reconstruction originating
from the thick target (see Methods for details). DALI2~\cite{takeuchi:2014:NIMA} surrounded MINOS to detect prompt de-excitation \gammarays with 
high efficiency. Secondary reaction species were subsequently identified with the ZeroDegree spectrometer from F8 to F11. An overview of the facility 
and the experimental set-up, including all the focal points, is provided in Fig.~\ref{fig:pid} together with obtained particle identification plots.

\section{\lowercase{\gammarays from the \ts{79}}C\lowercase{u\mbox{\boldmath{$(p,2p)$}}\ts{78}}N\lowercase{i reaction}} %% 36/40
Inclusive reaction cross sections for the production of \ts{78}Ni following the \ts{79}Cu$(p,2p)$\ts{78}Ni and \ts{80}Zn$(p,3p)$\ts{78}Ni reactions 
were 1.70(42) and 0.016(6)~mbarn, respectively, yielding 937 and 815 events. Energies of coincident prompt \gammarays were corrected for the Doppler 
shift in the spectra shown in Fig.~\ref{fig:gammaspectrum}. 
For the $(p,2p)$ reaction channel, the most intense \gammaray transition was observed at 2,600(33)~keV (error, standard deviation, s.d.) with a significance
level (S.L.) of 7.5 and tentatively assigned to the \etwopt decay of \ts{78}Ni. Four weaker transitions located at 583(10), 1,103(14), 1,540(25), and 
2,110(48)~keV were identified. 
All decay strengths, corrected for the \gammaray-energy dependent detection efficiency of DALI2, and confidence levels, are summarised in Extended 
Data Table~\ref{tab:conflevel} {and in Extended Data Fig.\ref{fig:confidence levels}}.
Sufficient statistics allowed to establish prompt coincidences
between the 2,600-keV transition and all the weaker transitions (Fig.~\ref{fig:gammaspectrum}b), as well as a possible coincidence between the 583-keV 
transition and the 2,110-keV transition (Extended Data Fig.~\ref{fig:extendedspectrum}b).
Conversely, no coincidence was observed between the 583- and 1,103-keV transitions (Extended Data Fig.~\ref{fig:extendedspectrum}b,c). They are, 
therefore, assigned to independently decay into the proposed \stwop state. On account of its correspondence to the theoretical descriptions discussed below, the 583-keV line
is tentatively assigned to the \efourpt transition. This leads to a ratio between the \etwop and \efourp of \eratior = 1.22, which 
is comparable to well-known doubly magic nuclei \ts{40}Ca~(1.35), \ts{48}Ca~(1.18), \ts{56}Ni~(1.45), \ts{132}Sn~(1.09),
and \ts{208}Pb~(1.06). 

\section{\lowercase{\gammarays from the \ts{80}}Z\lowercase{n\mbox{\boldmath{$(p,3p)$}}\ts{78}}N\lowercase{i reaction}} %% 36/40
Although similar amounts of $(p,2p)$ and $(p,3p)$ events were detected, the findings were largely different. 
A transition remained visible at 1,067(17) keV in the $(p,3p)$ spectrum (Fig.~\ref{fig:gammaspectrum}d), 
but no further prominent peak was observed in the energy range up to 2,600~keV, with transitions reported for the $(p,2p)$ channel possessing 
only S.L. around 1. A surprising additional strength above the \etwopt decay revealed a transition at 2,910(43)~keV with a S.L. of 3.5, which is 
either weakly or not populated in the $(p,2p)$ channel. 
It could not be interpreted as a decay into the \stwop state due to the low intensity of the peak at 2,600~keV.
Instead, it is ascribed to the decay of a second $2^{+}$ state to the $0^{+}$ ground state. This level placement is further corroborated by the  
spectrum for \gammaray detection multiplicities of $M_{\gamma} <4$ (Fig.~\ref{fig:gammaspectrum}c), which enhances the peak-to-total ratio of decays 
from low-lying levels, in this case the 2,910-keV transition. Applying similar arguments to the 1,067-keV transition, its intensity is too large 
to be identical with the 1,103-keV transition observed in the $(p,2p)$ channel, and is, therefore, placed as feeding the 2,910-keV level. 
Taking all these observations into account, the level scheme shown in Fig.~\ref{fig:levelscheme}b is proposed for \ts{78}Ni.

%\section{Discussion}
\section{D\lowercase{oubly magic \ts{78}}N\lowercase{i}} % 17/40
The \etwop along the chain of nickel isotopes are presented in Fig.~\ref{fig:levelscheme}a, which exhibit a local maximum at \ts{68}Ni that is 
attributed to the $N=40$ harmonic oscillator (HO) shell closure. This assessment is reinforced 
by its low {quadrupole} collectivity~\cite{sorlin:2002:PRL}. 
However, mass measurements identified it as a very localised 
feature~\cite{guenaut:2007:PRC} {and ref.~\cite{langanke:2003:PRC} pointed out that quadrupole collectivity is
dominated by neutron excitations.} 
Beyond the HO shell closure, {in the independent particle model} neutrons fill the $\nu 0g_{9/2}$ SPO with little impact on the \etwop 
until a steep rise is observed for our value at $N=50$. In fact, the 2.6~MeV excitation energy is essentially as high as the 
2.7~MeV for the doubly magic \ts{56}Ni~($Z=N=28$)\cite{ensdf}, thus providing first direct experimental evidence for a comparable
magic character. 
To gain further insight into the structural evolution of the neutron-rich nickel isotopes, their \etwop values were confronted with state-of-the-art 
theoretical calculations in Fig.~\ref{fig:levelscheme}a. Large-scale shell model (LSSM) 
calculations comprised the two shell-model Hamiltonians outlined in refs.~\cite{nowacki:2016:PRL,lenzi:2010:PRC} with a transition at $N=44$ from the 
LNPS to the PFSDG-U Hamiltonian. In particular, LSSM calculations for \ts{78}Ni included the full proton $pf$ and neutron $sdg$ HO shells into their
model space, which is crucial to pick up emerging quadrupole collectivity from quasi-degenerated neutron SPOs~\cite{nowacki:2016:PRL}.

The Monte Carlo Shell Model (MCSM)~\cite{shimizu:2012:PTEP} allows to incorporate more SPOs into the calculation. For the MCSM predictions presented
here, the A3DA-m Hamiltonian~\cite{tsunoda:2014:PRC} was employed for \ts{64--76}Ni, which encompasses the full $pf$ shell as well as the
$0g_{9/2}$ and $1d_{5/2}$ SPOs for protons and neutrons. To enable more detailed calculations for \ts{78}Ni, the A3DA-m Hamiltonian was extended to the
full proton and neutron $pf$ and $sdg$ shells. The mean field based quasi-particle random-phase approximation (QRPA)~\cite{peru:2014:EPJA} 
calculations implemented the Gogny D1M effective force \cite{goriely:2009:PRL}. 
Finally, we present {new} {\it ab initio} results based on the valence-space formulation~\cite{stroberg:2017:PRL} of the 
in-medium similarity renormalization 
group (IM-SRG)~\cite{hergert:2016:PR} {and show the coupled-cluster (CC) method~\cite{hagen:2016:PRL}}, both using two- and 
three-nucleon interactions derived from chiral effective field theory~\cite{epelbaum:2009:RMP}.
All theoretical calculations well describe the pattern of \etwop along the chain of nickel isotopes in Fig.~\ref{fig:levelscheme}a, notably the 
large enhancement at $N=50$, thus confirming \ts{78}Ni as a doubly magic nucleus.

\section{T\lowercase{wo different shapes emerge}} %37/40

Complete predictions for the low-lying level structure of \ts{78}Ni are presented in Fig.~\ref{fig:levelscheme}b next to
the proposed experimental level scheme. The LSSM and MCSM calculations analogously predict competing spherical (quadrupole deformation parameter, ${\beta \sim 0}$) 
and prolate deformed (${\beta \sim 0.3}$) intruder configurations, with
one discrepancy. While LSSM puts the deformed $2^{+}$ state at 2.88~MeV, and thus below the spherical $2^{+}$ state at 3.15~MeV, for MCSM the 
deformed $2^{+}$ state at 2.89~MeV lies above the spherical $2^{+}$ state at 2.57~MeV. Respective $4^{+}$ states are located approximately
0.5--0.7~MeV above the $2^{+}$ states, hence justifying the tentative spin assignment for the experimental level at 3.18~MeV. In addition, 
calculated deformed $0^{+}$ states are located approximately 0.25~MeV below their respective deformed $2^{+}$ states. A possible, unobserved transition 
from the deformed $2^{+}$ state
to the deformed $0^{+}$ state is expected to be several orders of magnitude weaker than direct decays to the ground state due to the large energy 
difference of the latter. It is further noted that restricting the MCSM calculations to the A3DA-m Hamiltonian~\cite{tsunoda:2014:PRC} puts 
the first $2^{+}$ state to 2.89~MeV and the second $2^{+}$ state to 4.72~MeV, strikingly demonstrating the necessity for the inclusion of the 
full neutron $sdg$ shell to properly characterise low-lying deformed configurations~\cite{nowacki:2016:PRL}.

%%%%%%%%%%%%%%%%%  
\begin{figure}[!bt]
  \centering
  \includegraphics[width=8.5cm]{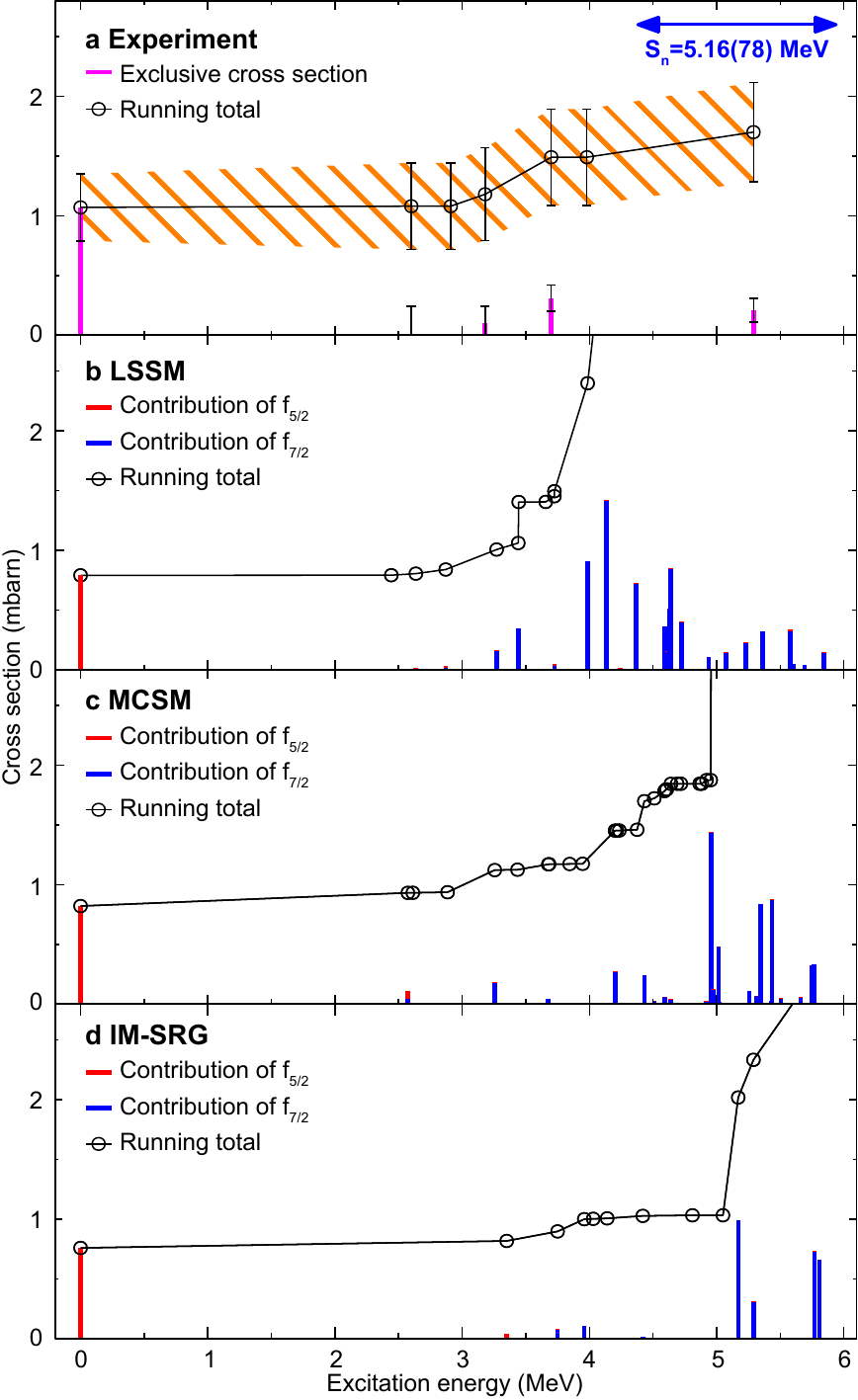}
  \caption{{\bf Experimental and calculated exclusive cross sections for the \ts{79}Cu\mbox{\boldmath{$(p,2p)$}}\ts{78}Ni reaction.}
    \textbf{a}, Measured exclusive cross sections (magenta vertical bars).
      The evaluated neutron separation energy, $S_n$, and its errors (1~s.d.) are indicated.
    \textbf{b}, Exclusive cross sections predicted by LSSM.
    \textbf{c}, Prediction by MCSM.
    \textbf{d}, Prediction by IM-SRG.
    In \textbf{b--d}, contributions of proton removal from $\pi0f_{5/2}$ and $\pi0f_{7/2}$ SPOs are distinguished by red and blue vertical bars, respectively.
    Integrated cross sections are shown as running total function of the excitation energy for the experimental and the theoretical predictions.
    {Contributions from the $\pi1p_{3/2}$ SPO amount to less than 0.05~mbarn and were omitted.} 
    Every circle along the running total corresponds to an observed or calculated level.
}
  \label{fig:cross_sections}% 
\end{figure} 
%%%%%%%%%%%%%%%%%

It is important to stress the structural differences between the spherical and deformed configurations, specifically the average number of particle-hole 
(p-h) excitations above $Z=28$ ($n^{\pi}_{\textrm{p-h}}$) and $N=50$ ($n^{\nu}_{\textrm{p-h}}$) for the $0^{+}$, $2^{+}$, and $4^{+}$ states. With MCSM, 
the numbers are ${0.4\lessapprox n^{\pi}_{\textrm{p-h}} \lessapprox 0.9}$ and ${0.7\lessapprox n^{\nu}_{\textrm{p-h}} \lessapprox 1.7}$ for the three spherical 
states, whilst
%range between 0.4 and 0.9 and the $n^{\nu}_{\textrm{p-h}}$ range between 0.7 and 1.7 for the three spherical states, while
the respective values are ${n^{\pi}_{\textrm{p-h}} \approx 2.5}$  
and ${n^{\nu}_{\textrm{p-h}} \approx 2.7}$ for the deformed states. Similar values are obtained with LSSM. Recollecting the contrasting levels
populated from $(p,2p)$ and $(p,3p)$ reactions, calculated spectroscopic factors, which quantify the overlap between final and initial state in
single nucleon removal reactions, can help unravel the nature of levels populated. 
Inspection of these spectroscopic
factors from \ts{79}Cu to final states in \ts{78}Ni with the LSSM and MCSM Hamiltonians shows spherical configurations are strongly favoured. 
Experimental exclusive cross sections to the individual levels are compared in Fig.~\ref{fig:cross_sections}
to calculated ones obtained  within the distorted-wave impulse approximation (DWIA) formalism~\cite{wakasa:2017:PPNP} folded with the 
shell-model spectroscopic factors (see Methods). {While the assumptions of the DWIA and shell-model spectroscopic factor 
calculations are not fully consistent, their combination provides a qualitative picture that can be compared to experiment.}
Note that for the removal of a single proton in the $\pi0f_{7/2}$ or $\pi0f_{5/2}$ SPO, calculated cross sections, which are weakly dependent on
projectile and excitation energy, are $\sim 1$~mbarn. The bulk of spectroscopic strength, which originates from the removal of a
$\pi0f_{7/2}$ proton, is calculated at energies between 4--5~MeV, {with an abrupt increase not observed experimentally.
However, this energy is} close or beyond the evaluated neutron separation
energy, $S_n$, of 5.16(78)~MeV~\cite{wang:2017:CPC}. Due to this large uncertainty of $S_n$, quantitative comparisons between the 
experimental and theoretical inclusive cross sections were not feasible, but it is noted that LSSM places the average of the distribution 
lower than MCSM.
Good agreement is observed for the ground state, which corresponds 
predominantly to a removal of a $\pi0f_{5/2}$ proton, and the strikingly low direct population of the observed low-lying levels, visualised by
means of the resembling running totals in Fig.~\ref{fig:cross_sections}. 
To date, no theoretical framework can predict microscopic $(p,3p)$ cross sections. It must be stressed, however, that
the calculated two-nucleon overlaps between the ground state in \ts{80}Zn and the excited states in \ts{78}Ni also favour the population of %predominantly produce 
spherical final states.

Nickel isotopes represent the neutron-rich frontier for \textit{ab initio} calculations.
For the IM-SRG results in Fig.~\ref{fig:levelscheme}b (for details, see Methods), we predict the \etwop of \ts{78}Ni to be 3.35~MeV and the \efourp 
at 3.75~MeV with a proton $pf$ and neutron $sdg$ SPO valence-space Hamiltonian. While this is several hundred keV higher than measured, it is 
nevertheless in good agreement with experimental trends across the chain and also predicts a sharp decrease in \etwop at \ts{80}Ni.
Average proton and neutron excitations for the \stwop are ${n^{\pi}_{\textrm{p-h}}=0.9}$ and ${n^{\nu}_{\textrm{p-h}}=1.3}$, analogous to the LSSM and MCSM, 
and also the exclusive cross sections (Fig.~\ref{fig:cross_sections}d) follow a similar trend. 
A stark variation is, however, found for the second $2^+$ state, which lies at 5.81~MeV and is of spherical nature.
This deficiency is not unexpected, as \textit{ab initio} methods are often built on truncations in allowed particle-hole excitations and thus 
fail to capture {very} collective features sufficiently. {In fact, the \etwop of \ts{78}Ni only 
varies about 100 keV when using several two- and three-nucleon interactions, so missing particle-hole excitations are likely the main 
uncertainty of the IM-SRG calculations. In particular, our IM-SRG results agree with the \textit{ab initio} CC predictions of
ref.~\cite{hagen:2016:PRL} at the level of singles and doubles correlations when using the same Hamiltonian.
When triples correlations are further included, the \etwop from CC for \ts{78}Ni is found to be in good agreement with the present 
measurement.}
%prediction~\cite{hagen:2016:PRL} is in good agreement with the experiment using the same two- and three-nucleon interaction.}}

% QRPA from sophie on Oct12
In the case of QRPA calculations, 65\% of the 2$^+_1$ wave function {of \ts{78}Ni} is composed of neutron excitations from 
the $\nu 0g_{9/2}$ to the $\nu 1d_{5/2}$ orbital, across the ${N=50}$ shell gap, while 28\% are proton excitations from the $\pi 0f_{7/2}$ to 
the $\pi 0f_{5/2}$ orbitals. In this approach, the neutron shell gap at $N=50$ is robust enough so that protons across the $Z=28$ shell gap 
significantly contribute to the excitation. Similarly, in the neighbouring even-even isotopes \ts{76}Ni and \ts{80}Ni, the $2^+_1$ excitation 
stems from neutron excitations in the $\nu 0g_{9/2}$ (80\% of the wave function) and in the $\nu 1d_{5/2}$ (72\% of the wave function with an 
occupancy of 0.3 neutrons) SPOs, respectively.

%%%
In conclusion, first direct experimental evidence for the preservation of the $Z=28$ and $N=50$ shell closures in \ts{78}Ni was provided
by employing a dedicated set-up to study extremely exotic nuclei via in-beam \gammaray spectroscopy. A low-lying, second 2$^+$ state conforms
with the notion of competing spherical and deformed configurations, 
which is to be verified by future measurements in \ts{76}Fe and \ts{74}Cr~\cite{nowacki:2016:PRL}.
Similarly, a breakdown of the proton $Z=28$ shell closure that favours prolate deformed ground states for heavier nickel isotopes is predicted, 
establishing \ts{78}Ni as doubly magic stronghold before deformation prevails in more exotic nuclei. 
Understanding these structures is crucial for r-process nucleosynthesis, as the onset of deformation may shift the drip line towards heavier isotopes.
Driving mechanisms for this evolution from spherical to deformed nuclei are yet to be fully
understood at the \textit{ab initio} level, prompting also future mass measurements to validate the predicted collapse of traditional shell closures.

%%%%%%%%%%%%%%%%%%%%%%%%%%%%%%%%%%%%%%
\makeatletter
\renewcommand\@biblabel[1]{#1.}
\makeatother
\newcommand{\bibitemnature}[1]{\addtocounter{enumii}{1}\bibitem[\number\value{enumii}]{#1}} %% To use numbers for methods summary also
\def\bibsection{\section{R\lowercase{eferences}}}
%%%%%%%%%%

%%%%%%%%%%%%%%%%%% 
%
\section{A\lowercase{cknowledgements}}
We express our gratitude to staff of the RIKEN Nishina Center accelerator complex for providing the stable and high-intensity uranium beam, and to the 
BigRIPS team for the smooth operation of the secondary beams.
We also thank for contributions by Mr.~Narumasa Miyauchi for providing us the 3D schematic figure of the RIBF facility shown in Fig.~\ref{fig:pid}a.
The development of MINOS and the core MINOS team have been supported by the European Research Council through the ERC Grant No.~MINOS-258567.
R.T. was supported by JSPS Grant-in-Aid for JSPS Research Fellow Grant Number JP14J08718.
A.O. was supported by JSPS long-term fellowship L-13520 at the RIKEN Nishina Center.
C.Sa. was supported by the IPA program at the RIKEN Nishina Center.
J.D.H acknowledges the support by National Research Council of Canada and NSERC.
This work was supported in part by the ERC Grant No.~307986 STRONGINT, the DFG under Grant SFB 1245, and the BMBF under Contract No.~05P18RDFN1.
The MCSM calculations were performed on K computer at RIKEN AICS (hp160211, hp170230, hp180179).
J.M., T.O., and Y.T. acknowledge the support from MEXT as "Priority Issue on post-K computer" 
(Elucidation of the Fundamental Laws and Evolution of the Universe) and JICFuS.
J.M. and K.O. were supported by Grant-in-Aid for Scientific Research JP18K03639 (J.M.) and JP16K05352 (K.O.).
A.Po. acknowledges the support by Mineco (Spain) grants FPA2014-57916 and Severo Ochoa Program SEV-2016-0597.
This work was supported in part by the DFG through the Cluster of Excellence PRISMA.
L.X.C. was supported by the Vietnam MOST through the Physics Development Program Grant No.~{\DJ}T{\DJ}LCN.25/18.
Zs.D., Z.K., and Zs.V. acknowledge the support from the GINOP-2.3.3-15-2016-00034 project.
M.L, C.L, and V.W. acknowledge the support from the German BMBF Grants No.~05P15RDNF1, No.~05P12RDNF8.

\section{A\lowercase{uthor Contributions}}
R.T. performed offline data analyses and GEANT4 simulations, and prepared the figures;
P.D., and A.O. designed the experiment;
R.T., C.Sa., P.D., A.O., J.D.H., J.M., F.N., K.O., T.O., A.S., and Y.T. wrote the manuscript;
R.T., C.Sa., P.D., A.O., G.A., D.C., F.C., A.C., A.D., J.-M.G, A.G., V.L., M.M., S.M., M.N., C.P,. A.Pe., E.C.P., J.-Y.R., Y.S., S.T., and H.W. 
were responsible for setting up the liquid hydrogen target and vertex reconstruction system, MINOS, and the $\gamma$-ray detector array, DALI2; 
R.T., C.Sa., H.B., D.C., A.C., and T.I. were responsible for the data acquisition system and analysis software;
R.T., C.Sa., P.D., A.O., G.A., H.B., D.C., F.C., A.C., A.D., J.-M.G, A.G., T.I., V.L., M.M., S.M., M.N., H.O., C.P., A.Pe., E.C.P., J.-Y.R., Y.S., S.T., H.W., F.B., L.X.C., Zs.D., S.F., F.G., A.G., K.H., Z.K., S.K., J.L., M.L., C.L., R.L., K.N., T.Mi., S.N., L.O., S.O., Z.P., E.S., C.Sh., P.-A.S., I.S., D.St., T.S., D.Su, Zs.V, V.W., J.W., and Z.Y.X. checked data accumulation online and maintained operation of the experiment;
P.D., A.O., K.Y., T.Mo., H.S., and T.U. supervised the participants;
F.N. and A.Po. performed the LSSM calculations;
T.O. and Y.T. performed the MCSM calculations;
S.P. performed the QRPA calculations;
J.D.H., J.M., A.S., J.S., and S.R.S. performed the IM-SRG calculations;
K.O. performed the DWIA calculations;
All authors discussed the results and commented on the manuscript.

\section{A\lowercase{dditional information}}
\noindent
\textbf{Reprints and permissions information} is available at www.nature.com/reprints.
\\
\textbf{Competing interests} The authors declare no competing interests.
\\
\textbf{Correspondence and requests for materials} should be addressed to P.D.

%%%%%%%%%%%%%%%%%

\cleardoublepage

%%%%%%%%%%%%%%%%%%

\section{M\lowercase{ethods}} %upto 3000words

\noindent\textbf{Beam production.}
The experiment was performed at the Radioactive Isotope Beam Factory, Japan, operated by the RIKEN Nishina Center and the Center for Nuclear Study 
of the University of Tokyo. A \ts{238}U primary beam with an intensity of $7.5\times10^{10}$~p.p.s. was accelerated by four coupled 
cyclotrons~\cite{okuno:2012:PTEP} to an energy of 345~\mevu and impinged on a 3-mm-thick beryllium production target, located at the F0 focus 
of the BigRIPS two-stage fragment separator~\cite{kubo:2012:PTEP} shown in Fig.~\ref{fig:pid}a.
The isotopes of interest were first purified with the BigRIPS fragment separator from focal plane F0 to F2 by their magnetic rigidity and energy loss.
Particles then traversed BigRIPS from F3 to F7 at approximately 60\% of the speed of light and were identified by time-of-flight, magnetic rigidity, 
and energy loss measurements on an event-by-event basis~\cite{fukuda:2013:NIMB,baba:2010:NIMA}. Secondary beam rates of 5~p.p.s.~for \ts{79}Cu 
and 290~p.p.s.~for \ts{80}Zn were directed onto the MINOS reaction target system.

\noindent\textbf{Secondary reaction.}
MINOS~\cite{obertelli:2014:EPJA} consisted of a tube of 102(1)-mm length and 38-mm inner diameter filled with liquid hydrogen coupled with a 
time projection chamber (TPC). Secondary reaction products were selected and identified with the ZeroDegree spectrometer following the same 
approach as for BigRIPS. Emitted prompt \gammarays were detected with DALI2~\cite{takeuchi:2014:NIMA}, which consisted of 186 NaI(Tl) scintillators 
arranged in a geometry that covered polar angles from 12 to 96 degrees surrounding the MINOS system and achieved a photo-peak efficiency of  26\% at 1~MeV. 
Figure~\ref{fig:pid}b shows the schematic view of DALI2 and MINOS.
Around 30\% energy loss inside the secondary target resulted in a highly reaction vertex position dependent velocity.
Therefore, an annular TPC surrounded the cryogenic target to determine the vertex position following $(p,2p)$ and $(p,3p)$ removal reactions.

\noindent\textbf{Data acquisition.}
The general data acquisition system at the RIBF facility~\cite{baba:2010:NIMA} was coupled with the dedicated MINOS 
electronics system~\cite{calvet:2014:IEEE,baron:2017:IEEE}
developed to readout the roughly 5,000~channels comprising the pad plane of the TPC.

\noindent\textbf{Vertex tracking and \gammaray energy reconstruction.}
Observed \gammaray energies in the laboratory system were reconstructed for their Doppler shift, as described in ref.~\cite{doornenbal:2012:PTEP}.
For this reconstruction, the vertex position was determined with the TPC by tracking the protons in the $(p,2p)$ and $(p,3p)$ knockout reactions 
of interest via an iterative tracking algorithm using a Hough transform method, detailed in ref.~\cite{santamaria:2018:NIMA}.
This process allowed a vertex resolution of less than $5~\textrm{mm}$ full width at half maximum.
The velocity at incident point was determined assuming energy loss inside the liquid hydrogen target according to the 
ATIMA code (https://web-docs.gsi.de/$\sim$weick/atima/).

Compton scattering of \gammarays resulted in partial energy deposition between neighbouring crystals of DALI2.
In the case \gammarays were observed in detectors within a 15-cm distance, the energy was summed and the detector with highest energy 
determined the position for the Doppler reconstruction.

The \gammaray detector response functions were simulated with the \textsc{geant4} framework~\cite{agostinelli:2003:NIMA} for 
fitting and comparison with the experimental data.

Stationary sources of \ts{60}Co, \ts{88}Y, and \ts{137}Cs provided \gammaray energies to calibrate the array.  

\noindent\textbf{Energy reconstruction of peaks.}
Five and six peak candidates were investigated in the $(p,2p)$ and $(p,3p)$ channels, respectively. The energy values and their corresponding
errors and intensities were determined by maximising the likelihood values of the fits {of 10~keV binned spectra} with all 
combinations of transition energies assuming a multivariate Gaussian probability density function. The S.L. for the existence of 
each peak was deduced from the probability value of the likelihood 
ratio between the null hypothesis and the alternative hypothesis. Obtained results are summarised in Extended Data Table~\ref{tab:conflevel}
{and in Extended Data Fig.\ref{fig:confidence levels}}.

%%%%%%%%%%%%%%%%%%%%%%%%%%%%%%%%%%%%%%%%%%%%%% Theory
%%%%%%%%%%LSSM from Nowacki
\noindent\textbf{LSSM calculation.}
Following ref.~\cite{nowacki:2016:PRL}, Large Scale Shell-Model Calculations were carried out in a valence space comprising the $pf$ shell orbitals 
for protons and the $sdg$ shell orbitals for neutrons, with an effective interaction based on realistic matrix elements for which the monopole part 
was adjusted to reproduce the experimental evolution of the regulating gaps in the model space.
Besides the convincing spectroscopy exhibited in the comparison panels of Fig.~\ref{fig:levelscheme}, the spectroscopic qualities of this interaction 
was already assessed in the recent studies in the region~\cite{welker:2017:PRL, wraith:2017:PLB, groote:2017:PRC, shand:2017:PLB, cortes:2018:PRC}. 
In order to generate the theoretical spectroscopic strength of one proton removal from $^{79}$Cu to final states in $^{78}$Ni, the Lanczos Strength 
Function Method~\cite{Whitehead,caurier:2005:RMP} was applied to LSSM calculations. 
In practice, the removal operator was applied on the initial $^{79}$Cu ground state to generate the sum rule 
state $\Sigma = \tilde{a}_j \left\vert \frac {5}{2}^- \right\rangle$.
This non physical state carried the entire one proton removal spectroscopic strength and was used as the initial state for a Lanczos diagonalisation 
procedure. At the end of the procedure, the matrix elements were simply the components of $\Sigma$ in the Lanczos eigenstates basis.
Even if the diagonalisation was not fully complete; this procedure converged rapidly to an approximate strength distribution. 
To handle more easily the large dimension 
of the involved basis without losing any physics insight, the proton and neutron gaps were slightly reduced to project the results of 
ref.~\cite{nowacki:2016:PRL} into a smaller valence space containing all excitations across the $Z=28$ and $N=50$ up to 7 particle-hole 
excitations. We checked that the orbital occupancies for $^{78}$Ni low-lying states and $^{79}$Cu ground-state wave functions remained identical to 
the ones obtained in the configuration space of ref.~\cite{nowacki:2016:PRL}.

\noindent\textbf{MCSM calculation.}
The MCSM calculation is one of the most advanced computational methods that can be applied for nuclear many-body systems.
This work represents the MCSM calculations performed on the K computer at RIKEN AICS, Japan.
Exploiting the advantages of quantum Monte Carlo variational and matrix-diagonalisation methods, this approach circumvents the diagonalisation 
of a $>5 \times 10^{20}$-dimensional Hamiltonian matrix. Using the doubly magic \ts{40}Ca nucleus as an inert core, 8 protons and up to 32 neutrons 
were left to actively interact in a much larger model space as compared to conventional configuration interaction calculations.
The A3DA-m interaction~\cite{tsunoda:2014:PRC} in the model space of 6~SPOs ($pf$ shell, $0g_{9/2}$ and $1d_{5/2}$ orbitals) both for protons and 
neutrons was used for \ts{64-76}Ni. This interaction was extended to a model space of 9~SPOs ($pf$ and $sdg$ shells) for \ts{78,80}Ni where 
upper SPOs in the $sdg$ shell become more important. The MCSM eigenstate is represented as a superposition of MCSM basis vectors with the appropriate 
projection onto spin and parity. Each MCSM basis vector is a Slater determinant formed by mixed single-particle states, where the mixing 
amplitudes are optimised by quantum Monte Carlo and variational methods.
This Slater determinant has intrinsic quadrupole moments, which can be expressed in terms of a set of $\beta_2$ and $\gamma$ deformation parameters.
This property can be used to study the intrinsic shape of the calculated state.

The MCSM calculations with the A3DA-m interaction reveal properties of Ni and neighbouring isotopes. 
Shape evolution of nickel isotopes including shape coexistence has been described theoretically~\cite{tsunoda:2014:PRC} and the calculations 
have been compared with experiments for \ts{66}Ni~\cite{leoni:2017:PRL}, \ts{68}Ni~\cite{suchyta:2014:PRC,flavigny:2015:PRC}, 
and \ts{70}Ni~\cite{chiara:2015:PRC,morales:2017:PLB}. Nuclei near $N=50$ have been described well by the calculations for 
\ts{77}Cu~\cite{sahin:2017:PRL} and \ts{79}Cu~\cite{olivier:2017:PRL}. 

\noindent\textbf{In-medium similarity renormalization group.}
The in-medium similarity renormalization group (IM-SRG)~\cite{hergert:2016:PR,tsukiyama:2011:PRL} transforms the many-body 
Hamiltonian, ${H}$, to a diagonal or block-diagonal form, $\tilde{H}={U}{H}{U}^{\dagger}$, via a unitary transformation, ${U}=e^{{\Omega}}$, 
where the anti-Hermitian generator, ${\Omega}$, builds in the off-diagonal correlations from the original $H$~\cite{morris:2015:PRC}.
The IM-SRG starts from a single-reference ground-state configuration, $| \Phi_0 \rangle$ (such as Hartree-Fock), at the flow parameter, $s=0$, 
and connects this state to the fully correlated ground state, $|\Psi_0 \rangle$, as $s \to \infty$, where the unitary transformation 
is also built up as a continuous sequence of unitary transformations, $U(s)$. With no truncations in the IM-SRG flow equations, this 
gives the exact ground-state energy, but in the IM-SRG(2) approximation used here, all operators are truncated at the two-body level.
In the valence-space IM-SRG formulation~\cite{stroberg:2017:PRL,tsukiyama:2012:PRC,bogner:2014:PRL,stroberg:2016:PRC}, the unitary 
transformation is constructed to in addition decouple the valence space, leading to a valence-space Hamiltonian, which is subsequently 
diagonalised within the valence space using large-scale shell-model methods. In this work, this encompasses {a \ts{60}Ca 
core and a Hamiltonian composed of} the proton $pf$ and neutron $sdg$ SPOs. 

Here, we used a particular set of two- and three-nucleon interactions (labelled 1.8/2.0 (EM)~\cite{hebeler:2011:PRC}), which was fit in 
few-body systems, but predicts ground state energies very well to $^{100}$Sn~\cite{morris:2018:PRL}.
{The calculations adequately reproduce the \etwop trend from \ts{70}Ni to } the steep rise in excitation energy for \ts{78}Ni.
In addition, the 1.8/2.0 (EM) interaction within the IM-SRG framework systematically predicts experimental binding energies and \etwop of 
lighter nuclei up to the nickel isotopes, including high \etwop values at known shell closures~\cite{simonis:2017:PRC}. {As 
for LSSM, spectroscopic factors  were obtained with the Lanczos method~\cite{Whitehead}.}

{
The shell model diagonalisation of nickel isotopes up to \ts{72}Ni is exact in the valence space. Due to exceeding dimensions, for 
\ts{74}Ni and \ts{78}Ni, the number of particle-hole excitations across the proton $f_{7/2}$ orbital ($Z=28$) and the neutron $g_{9/2}$ 
orbital is limited to 7$p$-7$h$. Similarly, for \ts{76}Ni and \ts{80}Ni up to 6$p$-6$h$ and 5$p$-5$h$ excitations were included, 
respectively. The convergence of the results in terms of $np$-$nh$ excitations suggests uncertainties of about 10~keV for \ts{74,76}Ni 
and 50~keV for \ts{78,80}Ni. The spectroscopic factors shown if Fig.~\ref{fig:cross_sections}d are calculated diagonalising \ts{79}Cu 
including up to 6$p$-6$h$ excitations. At this level, the convergence of the largest spectroscopic factors is better than 5~\%. 
Thus, these uncertainties are smaller than the uncertainties from the input two- and three-nucleon interactions as well as the IM-SRG 
truncations.}

\noindent\textbf{QRPA calculation.}
In the presented QRPA approach, the second-order matrix was fully diagonalised \cite{peru:2008:PRC,peru:2014:EPJA}, in agreement with the 
variational principle applied in Hartree-Fock-Bogoliubov (HFB) mean field calculations. Eigenvalues and eigenvectors provided excitation energies 
and wave functions for all vibrational excited states described as coherent states of two quasi-particle excitations above the ground state.
Transition probabilities from the ground state to all described excited states could be calculated for all parities and multipolarities, and 
for any intrinsic deformation of the ground state. Not only spherical, but also axially-symmetric deformed shapes were considered. QRPA and 
underlying HFB equations were solved in a finite HO basis with cylindrical coordinates including eleven major shells.
In order to preserve symmetry restoration, the same curvature of the HO was given for the both radial and symmetric axis directions.
This curvature was chosen for each calculated nucleus in order to minimise the HFB binding energy: The oscillator parameter, $\hbar \omega$, 
as defined in a previous study\cite{gaudefroy:2005:PRC}, was found to be 9.9~MeV for $^{78}$Ni.

All HFB quasi-particle states were used to generate the two-quasi-particle excitation set. This means that no additional cut in energy, 
occupation probabilities nor isospin was introduced. The same effective interaction, namely the D1M parameter set \cite{goriely:2009:PRL} 
of the Gogny analytical form, was used to solve both the HFB and the QRPA equations in all $ph$, $pp$, and $hh$ channels.
Since the D1M effective interaction has been fit without exchange and pairing parts of the Coulomb interaction, only the direct term of the 
Coulomb field is included in the HFB as well as in QRPA matrix elements. As concluded in a previous study\cite{anguiano:2001:NPA}, this term 
should have an impact only in case of large proton pairing energy, which is not the case here where the $Z=28$ is a closed proton shell.
The two-body centre-of-mass correction term was also neglected in the present study, as its impact is expected to be very small.

{The QRPA approach provides good agreement with the experimental \etwop, as shown in Fig.~\ref{fig:levelscheme}. Note that 
the \etwop of \ts{64}Ni has been omitted, as its spherical shape is degenerated with
an oblate deformation (see Fig.~42 of ref.~\cite{peru:2014:EPJA}) and therefore not suitable for QRPA calculations.}

\noindent\textbf{DWIA calculation.}
DWIA describes a $(p,2p)$ process as a proton-proton ($pp$) scattering inside a
target nucleus, $A$. The transition amplitude, $T$, whose absolute square is
proportional to the reaction probability, is given by,%
\[
T=\left\langle \chi_{1}\chi_{2}\left\vert v\right\vert \chi_{0}\phi
_{N}\right\rangle ,
\]
where $\chi$ represents the scattering waves of the incoming nucleon ($\chi_{0}$)
and the outgoing two nucleons ($\chi_{1}$ and $\chi_{2}$), $\phi_{N}$ is a
bound-state wave function of a nucleon to be knocked out, and $v$ is the $pp$
interaction that causes the ($p,2p$) transition. In the plane-wave limit, $T$
becomes a product of the Fourier transforms of $v$ and $\phi_{N},$ which
allows one to interpret a ($p,2p$) cross section as a snapshot of the momentum
distribution of the struck proton inside $A$. In reality, distortion effects
such as deflection and absorption exist, which are well under control by using
distorted waves for the incoming proton and outgoing protons. ($p,2p$)
reactions have been established as an alternative to ($e,e^{\prime}p$)
reactions. More details can be found in ref.~\cite{wakasa:2017:PPNP}.

In the present work, we constructed each $\chi$ by a microscopic $G$-matrix
folding model calculation, that is, the $pp$ interaction in infinite nuclear
matter was folded by a nuclear density of $A$. We adopted the Melbourne
$G$-matrix \cite{Amos00} for the former and the Bohr-Mottelson (BM) density \cite{BM69}
for the latter. The microscopic $p$-$A$ potential was shown to describe proton
scattering observables for various nuclei in a very wide range of energies~\cite{Amos00}.
The BM parameter was used also for $\phi_{N}$. The correction to the nonlocality of
$\chi$ and $\phi_{N}$ was accounted for by the prescription proposed by Perey
and Buck~\cite{perey:1962:NP}. As for the transition interaction $v$, the Franey-Love
parametrisation~\cite{franey:1985:PRC} was employed. All the central, spin-orbit, and tensor parts
were included.
In Fig.~\ref{fig:cross_sections}, spectroscopic strengths from more deeply bound SPOs are not illustrated as they are negligibly small.

%%%%
\section{D\lowercase{ata availability}}
All of the relevant data that support the findings of this study are available from the corresponding author upon reasonable request.

%%%%%%%%%%%%%%%%%%%%%%% Citation for Method section. No limit.

%%%%%%%%%%

%%%%%%%%%%%%%%%%%%%%%%%%%%%%%%%%%%%%%%%%%%%%%%%
%% Extended data %%

%\cleardoublepage
\renewcommand{\figurename}{\textbf{Extended Data Fig.}}
\renewcommand{\tablename}{\textbf{Extended Data Table}}
\setcounter{figure}{0}
\setcounter{table}{0}

%%%%%%%%%%%%%%%%%% 
\begin{figure*}[pt]
\centering
        \includegraphics[width=18.3cm]{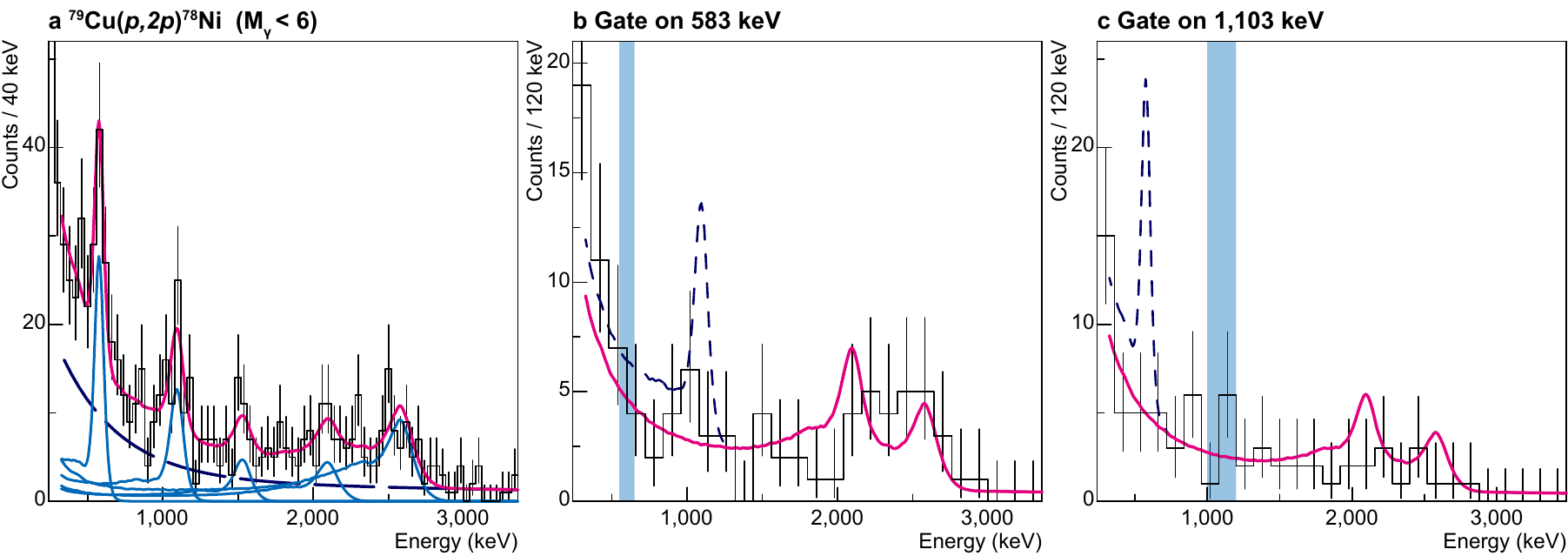}
\caption{\textbf{Energy spectra of prompt \gammaray coincidences with \ts{79}Cu\mbox{\boldmath{$(p,2p)$}}\ts{78}Ni reactions}.
  \textbf{a}, Same as Fig.~\ref{fig:gammaspectrum}a with a binning condition of 40~keV, allowing to resolve the transition at 583~keV.
  \textbf{b}, \gammaray spectrum in coincidence with the 583-keV transition.
              Expected intensities for coincidences with the 583-keV transition are indicated by the simulated lineshapes, with and without the 1,103-keV transition shown by blue dashed and magenta solid lines, respectively.
              It reveals of no coincidence between the 583- and 1,103-keV transitions.
  \textbf{c}, \gammaray spectrum in coincidence with the 1,103-keV transition.
              The hypothesis of no coincidence between the 583- and 1,103-keV transitions is corroborated.
              Coincidence ranges are illustrated by the hatched area in the respective spectra in \textbf{b} and \textbf{c}.
    }%    
\label{fig:extendedspectrum}% 
\end{figure*}

%%%%%%%%%%%%%%%%%% 
%\clearpage
\begin{table*}[pt]
\centering
\caption{\textbf{Observed \gammaray transition energies, relative intensities, and significance levels for the \ts{79}Cu\mbox{\boldmath{$(p,2p)$}}\ts{78}Ni and \ts{80}Zn\mbox{\boldmath{$(p,3p)$}}\ts{78}Ni reaction channels.}
}
        \includegraphics[width=8.9cm]{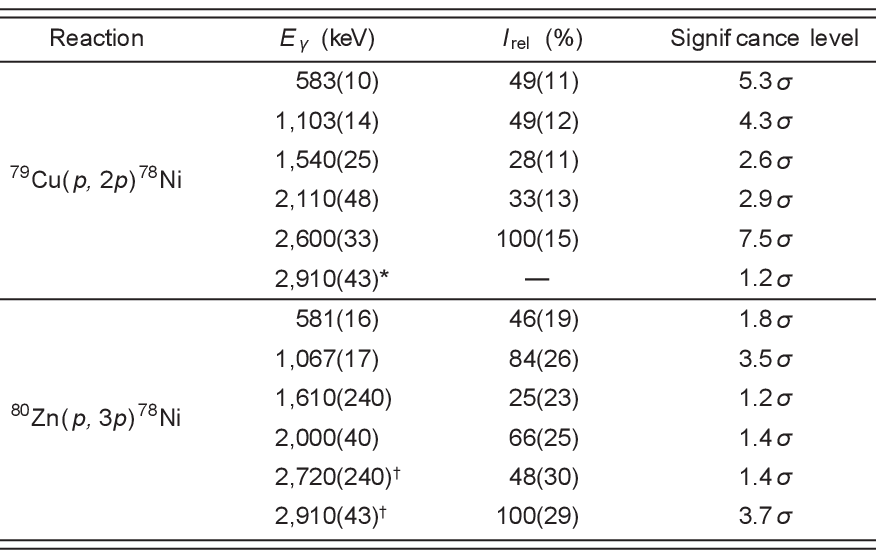}\\
%%% To be 7pt helvetica. Later, it should be converted into EPS format
%\textsf{\fontsize{7pt}{0.4cm}\selectfont
%\begin{minipage}[b]{8.9cm}
%\begin{ruledtabular}
%\begin{tabular}{clcc}
%%\hline
%Reaction& $E_\gamma$ (keV) & $I_{\mathrm{rel}}$ (\%) & Significance level \\\hline
%\multirow{6}{*}{\ts{79}Cu$(p,2p)$\ts{78}Ni}
%&\phantom{0,}583(10)&\phantom{0}49(11)&5.3$\sigma$\\
%&1,103(14)      & \phantom{0}49(12)	&4.3$\sigma$\\
%&1,540(25)      & \phantom{0}28(11)	&2.6$\sigma$\\
%&2,110(48)	& \phantom{0}33(13)	&2.9$\sigma$\\
%&2,600(33)	& 100(15)	&7.5$\sigma$\\ 
%&2,910(43)${}^*$&---&1.2$\sigma$\\
%\hline
%\multirow{6}{*}{\ts{80}Zn$(p,3p)$\ts{78}Ni}
%&\phantom{0,}581(16)& \phantom{0}46(20)	&1.6$\sigma$\\
%&1,067(17) 	& \phantom{0}82(27)	&3.5$\sigma$\\
%&1,670(240)	& \phantom{0}19(24)	&0.7$\sigma$\\
%&2,000(40) 	& \phantom{0}65(26)	&2.4$\sigma$\\
%&2,720(240)${}^{\dag}$	& \phantom{0}48(30)	&1.4$\sigma$\\
%&2,910(43)${}^{\dag}$ 	& 100(30)	&3.5$\sigma$\\
%\end{tabular}
%\end{ruledtabular} 
%\end{minipage}
%}
\begin{flushleft}
{De-excitation energies, $E_\gamma$, with statistical errors (1 s.d.) determined by maximising likelihoods in probability density functions and the 
relative intensity, $I_{\mathrm{rel}}$, to the most intense transition for each reaction, are listed. 
{Shown are the significance levels for \gammaray detection multiplicity $M_{\gamma} < 6$}
\\
${}^*${The significance level was tested with the obtained energy of 2,910~keV from the $(p,3p)$ reaction.}
\\
${}^{\dag}${The 2,600-keV transition observed in the $(p,2p)$ reaction was not observed distinctly in the $(p,3p)$ reaction, resulting 
in a {likelihood maximum} at 2,720(240)~keV.
The energy and the significance level of the transition at 2,910(43)~keV were determined by fixing the neighbouring transition to 2,600~keV.}
}
\end{flushleft}
\label{tab:conflevel}% 
\end{table*}
%%%%%%%%%%%%%%%%%

%%%%%%%%%%%%%%%%%% 
%\clearpage
\begin{figure*}[pt]
\centering
        \includegraphics[width=17.3cm]{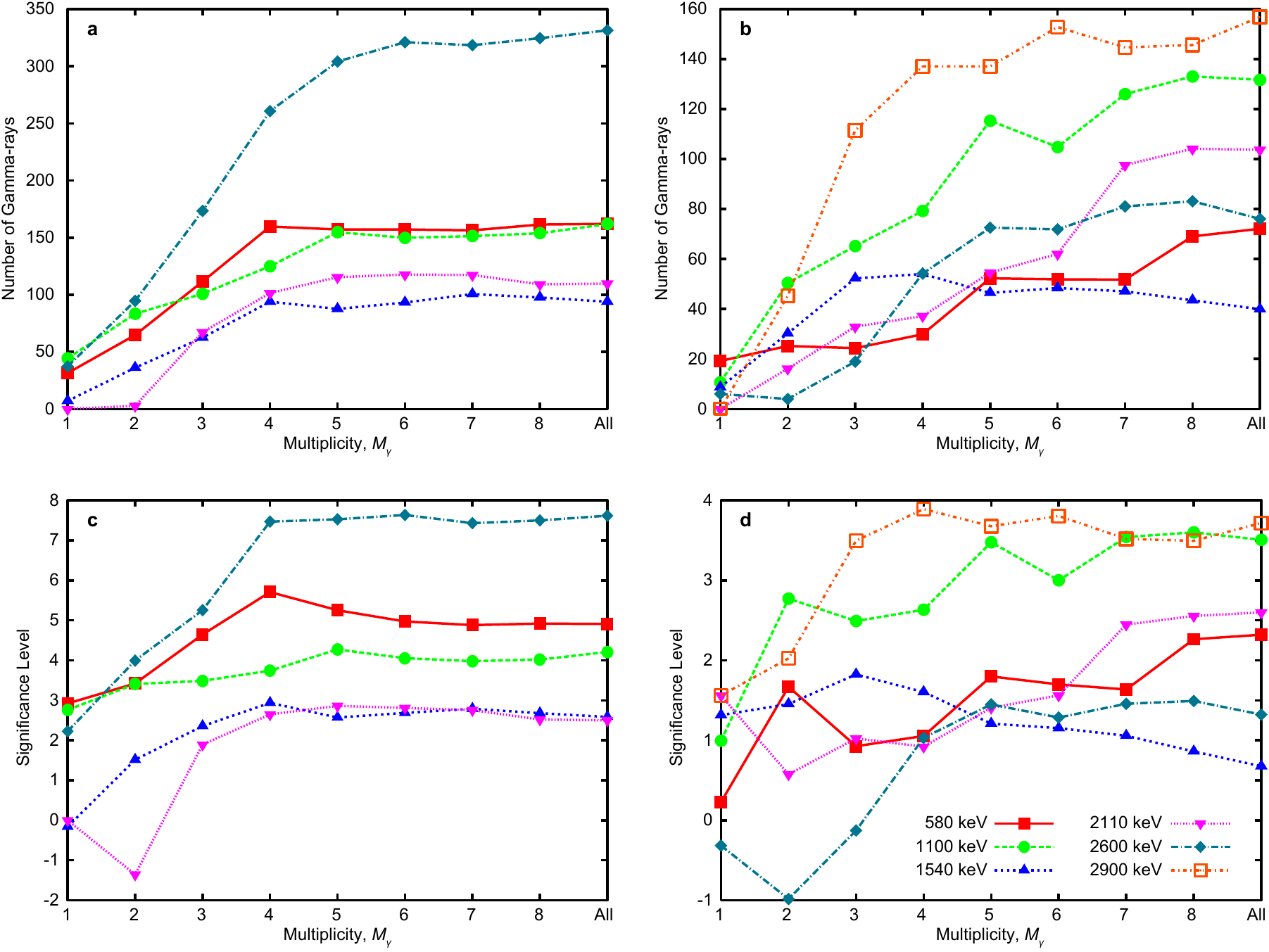}
\caption{\textbf{Evolution of peak significance and fitted intensities as function of \gammaray multiplicity}.
  \textbf{a,b}, Number of emitted \gammarays obtained for fitted findividual transition   
  \textbf{c,d}, Significance levels of individual transitions. The left side corresponds to the \ts{79}Cu$(p,2p)$\ts{78}Ni reaction
  and the right side to the \ts{80}Zn$(p,3p)$\ts{78}Ni reaction.
    }%    
\label{fig:confidence levels}% 
\end{figure*}

\end{document}